\pdfoutput=1
\documentclass[%
 reprint,
 amsmath,amssymb,
 aps,
]{revtex4-2}

\usepackage{graphicx}
\usepackage{dcolumn}
\usepackage{bm}
\usepackage{mathtools}
\usepackage{hyperref}

\usepackage{xcolor}

\begin{document}

\preprint{APS/123-QED}

\title{Improving the scalability of Gaussian-process error marginalization in gravitational-wave inference}

\author{Miaoxin Liu}
 \affiliation{Department of Physics, National University of Singapore, Singapore 117551}

\author{Xiao-Dong Li}
\affiliation{%
School of Physics and Astronomy, Sun Yat-sen University Zhuhai Campus, 2 Daxue Road, Tangjia, Zhuhai 519082, P.R. China\\
 CSST Science Center for Guangdong-Hong Kong-Macau Great Bay Area, Zhuhai 519082, P.R. China 
}%

\author{Alvin J. K. Chua}%
\email{alvincjk@nus.edu.sg}
\affiliation{Department of Physics, National University of Singapore, Singapore 117551\\
 Department of Mathematics, National University of Singapore, Singapore 119076 
}%


\date{\today}

\begin{abstract}
The accuracy of Bayesian inference can be negatively affected by the use of inaccurate forward models. In the case of gravitational-wave inference, accurate but computationally expensive waveform models are sometimes substituted with faster but approximate ones. The model error introduced by this substitution can be mitigated in various ways, one of which is by interpolating and marginalizing over the error using Gaussian process regression. However, the use of Gaussian process regression is limited by the curse of dimensionality, which makes it less effective for analyzing higher-dimensional parameter spaces and longer signal durations. In this work, to address this limitation, we focus on gravitational-wave signals from extreme-mass-ratio inspirals as an example, and propose several significant improvements to the base method: an improved prescription for constructing the training set, GPU-accelerated training algorithms, and a new likelihood that better adapts the base method to the presence of detector noise. Our results suggest that the new method is more viable for the analysis of realistic gravitational-wave data.


\end{abstract}

\maketitle


\section{\label{sec:level1} Introduction}

The field of gravitational-wave (GW) astronomy has witnessed remarkable progress so far, with the detection of approximately 90 compact binary coalescences (stellar-mass binary mergers) by the LIGO-Virgo-KAGRA Collaboration
\cite{abbott2021gwtc,LIGOScientific:2021psn}. Future space-based GW detectors operating in the millihertz frequency band, namely LISA \cite{DEA2017}, TianQin \cite{mei2021tianqin,luo2016tianqin}, and Taiji \cite{Ruan:2020smc}, will lead to the discovery of new kinds of sources such as binary white dwarfs \cite{huang2020science,Korol:2017qcx}, massive binary black-hole mergers \cite{wang2019science,Klein:2015hvg}, stellar-mass binary inspirals \cite{liu2020science,Sesana:2016ljz}, and extreme-mass-ratio inspirals (EMRIs) \cite{Babak:2017tow,Fan:2020zhy,BEA2017,BEA2019,GTV2010}. Gravitational waves generated by all of these extreme astronomical events carry unique information, providing novel insights into the physics and astronomy of such phenomena.

To achieve scientific goals in GW astronomy, it is essential to identify and characterize GW signals within a noisy data stream. The characterization process involves the inference of astrophysical parameters based on a certain GW source model. The accuracy of parameter estimation is constrained by two factors: the statistical error caused by the noise and the theoretical error due to the use of an inaccurate waveform model. It is known that, for both ground-based \cite{Canitrot:2001hc} and space-based GW detectors \cite{Cutler:2007mi}, the statistical error decreases as the signal-to-noise ratio (SNR) increases, while the theoretical error remains constant; this may lead to the exclusion of the true parameter values with high statistical significance.

In GW data analysis, the \emph{deliberate} incurrence of theoretical error is a common scenario, as it occurs whenever fast approximate models are used in lieu of more accurate but computationally costly models/simulations (e.g., waveforms from numerical-relativity simulations \cite{Mroue:2013xna,LIGOScientific:2014oec}).
To account for the presence of (known) theoretical error, Gaussian-process regression (GPR) \cite{M2003,RW2006}, a machine-learning technique, has been proposed as a method for interpolating and marginalizing over such error \cite{MG2014}. The method fits a Gaussian process to a small set of precomputed waveform differences between an accurate fiducial model and an approximate one. This process then serves as a prior distribution for the waveform difference, and can be marginalized over in the standard Bayesian likelihood with the approximate model. The GPR marginalized likelihood, which is informed by accurate waveforms, corrects the search under approximate templates and accounts for any residual model inaccuracy with (generally) more conservative error estimates. This method has since been applied in follow-up studies \cite{MBCG2016,Chua:2016jnd,Moore:2016wcn,Chua:2019wgs}.

Previous research has demonstrated the potential of the GPR marginalized-likelihood method to mitigate theoretical error in low-dimensional cases \cite{MG2014,MBCG2016,Chua:2019wgs}. However, the curse of dimensionality is a major challenge that hinders the use of GPR even in general applications. The number of training points required to cover a parameter space typically increases exponentially with its dimensionality, while the computational complexity of GPR increases cubically with the size of the training set. In the marginalized-likelihood method, this not only slows down the offline training phase but also the online evaluation phase, negating the speed advantage of using approximate templates.

In this study, we propose multiple improvements to the base GPR method that better adapt it to high-dimensional cases; the most notable is the use of Fisher-information-based Latin hypercube sampling (LHS) to generate a more informative training set with fewer points. We illustrate the efficacy of our approach by applying it to EMRI parameter estimation with a representative ``accurate'' signal model \cite{Katz:2021yft} and an artificially constructed ``approximate'' template model. Even though accurate next-generation EMRI models \cite{Pound:2019lzj,Warburton:2021kwk,Wardell:2021fyy} will not be significantly more costly than existing approximate ones (due to recent computational developments \cite{Katz:2021yft,Chua:2020stf,speri}), we choose EMRIs as our example here because the intrinsic information content and computational complexity of their waveforms epitomize most of the difficulties that inhibit the use of the base GPR method.

The remainder of the paper is organized as follows. Sec.~\ref{subsec:marginalised_likelihood} provides a brief overview of the marginalized-likelihood method, while Sec.~\ref{subsec:GPR} introduces the technique of GPR in the context of waveform interpolation. The training of the GPR model using a precomputed set of waveform differences is discussed in Sec.~\ref{subsec:GPR training}, while Sec.~\ref{subsec:Fisher coordinate} reviews the construction of the training set as described in previous studies. These sections provide important background information for understanding the application of the GPR method in GW analysis.

In Sec.~\ref{sec:improvements}, we describe and demonstrate our proposed improvements to the base GPR method. We introduce a parameter re-scaling strategy in Sec.~\ref{subsec:re-scaling}, and discuss the new hyperparameters to be trained in Sec.~\ref{subsec:training_parameters}. We also compare the LHS training set construction method to the old method in Sec.~\ref{subsec:LHS}, concluding that the former contains more information than the latter with the same density. We describe computational enhancements to training in Sec.~\ref{subsec:efficiency}. We present a new form of the GPR marginalized likelihood that properly treats the presence of detector noise in Sec.~\ref{subsec:noise likelihood}, and discuss an iterative approach to the GPR method in Sec.~\ref{subsec:iterative GPR}. Our model is tested on data with simulated detector noise in Sec.~\ref{subsec:LISA}. Finally, we summarize the new techniques implemented in this work and propose possible computational strategies for further extensions in Sec.~\ref{sec:conclusion}.


\section{Background: The base GPR method}\label{sec:marginalized likelihood}
\subsection{Marginalized likelihood}\label{subsec:marginalised_likelihood}
For a two-channel GW detector, the single source data can be expressed as
\begin{eqnarray}
x(t) = s(t)+n(t),
\label{eq:one}
\end{eqnarray}
where $s\equiv(s_\mathrm{I},s_\mathrm{II})$ is the source signal and $n\equiv(n_\mathrm{I},n_\mathrm{II})$ is the detector noise. In the standard matched-filtering framework, the data is compared against waveform templates $h\equiv(h_\mathrm{I},h_\mathrm{II})$ that are parametrized by some astrophysical parameters $\boldsymbol{\theta}$, while the detector noise is treated as a Gaussian and stationary stochastic process.

The Bayesian likelihood of the source parameters is thus \cite{CF1994}
\begin{equation}\label{eq:standard_likelihood}
L\propto\exp{\left(-\frac{1}{2}\langle x-h|x-h\rangle\right)},
\end{equation}
where the noise-weighted inner product $\langle\cdot|\cdot\rangle$ on the space of finite-length time series is given by
\begin{equation}\label{eq:inner_product}
\langle a|b\rangle=4\,\mathrm{Re}\sum_{f>0}^{f_N}\mathrm{d}f\sum_{\chi=\mathrm{I},\mathrm{II}}\frac{\tilde{a}_\chi^*(f)\tilde{b}_\chi(f)}{S_{n,\chi}(f)},
\end{equation}
with overtildes denoting discrete Fourier transforms, $f_N$ denoting the Nyquist frequency, and $S_{n,\chi}$ denoting the one-sided power spectral density of the channel noise $n_\chi$. The optimal SNR of a waveform template $h$ is given in terms of this inner product as $\sqrt{\langle h|h\rangle}$, while the overlap between two templates is defined as $\langle h_1|h_2\rangle/\sqrt{\langle h_1|h_1\rangle\langle h_2|h_2\rangle}$.

If an accurate template model $h_\mathrm{acc}(\boldsymbol{\theta})$ is used for the analysis, then the source signal $s$ is well described by the model at the actual parameter values $\boldsymbol{\theta}_\mathrm{true}$, i.e., $s=h(\boldsymbol{\theta}_\mathrm{true})$. The maximum-likelihood estimate $\boldsymbol{\theta}_\mathrm{ML}$ does not generally equal $\boldsymbol{\theta}_\mathrm{true}$ due to the presence of detector noise, but the parameter error $\boldsymbol{\theta}_\epsilon=\boldsymbol{\theta}_\mathrm{ML}-\boldsymbol{\theta}_\mathrm{true}$ is purely statistical in nature as it arises only from $n$, and is thus fully described by the posterior. On the other hand, if an approximate template model $h_\mathrm{app}(\boldsymbol{\theta})$ is used for the analysis, then the parameter error now contains an additional contribution from the difference $h_\epsilon=h_\mathrm{app}-h_\mathrm{acc}$. For high-SNR sources, this theoretical-error term may exceed the statistical uncertainties described by the posterior, and thus become the limiting factor in obtaining accurate parameter estimates \cite{CV2007}.

The bias from theoretical error can be mitigated by specifying a suitable prior probability distribution $p(h_\epsilon)$ for the waveform difference $h_\epsilon$, then marginalizing over $h_\epsilon$ in Eq.~\eqref{eq:standard_likelihood}. This ``marginalized likelihood'' is given by
\begin{equation}\label{eq:marginalisation}
\mathcal{L}\propto\int_W\mathrm{D}h_\epsilon\,p(h_\epsilon)L_\mathrm{acc},
\end{equation}
where $L_\mathrm{acc}$ is Eq.~\eqref{eq:standard_likelihood} with $h=h_\mathrm{acc}=h_\mathrm{app}-h_\epsilon$, and $W$ is the space of waveform differences. In \cite{MG2014}, Moore \& Gair proposed using GPR to define a Gaussian prior distribution, thus allowing the above integral to be analytically approximated (since $L_\mathrm{acc}$ is also formally Gaussian).

\subsection{Gaussian process regression}\label{subsec:GPR}
In the GPR method, $h_\epsilon\in W$ may be modeled as a Gaussian process over the parameter space $\Theta$:
\begin{equation}\label{eq:GP}
h_\epsilon(\boldsymbol{\theta})\sim\mathcal{GP}(\bar{h}_\epsilon,k),
\end{equation}
where $\bar{h}_\epsilon$ is the (vector-valued) mean of the process, and $k(\boldsymbol{\theta},\boldsymbol{\theta}')$ is the covariance function of the process. Then the set of waveform differences $\{h_\epsilon(\boldsymbol{\theta}_i)\in W\,|\,i=1,2,...,N\}$ corresponding to a small training set of parameter points $\{\boldsymbol{\theta}_i\in\Theta\,|\,i=1,2,...,N\}$ has a Gaussian probability distribution $\mathcal{N}(\bar{h}_\epsilon,\mathbf{K})$ on $W^N$ \cite{Chua:2019wgs}:
\begin{equation}\label{eq:PDF}
p([h_\epsilon(\boldsymbol{\theta}_i)])=\frac{1}{(2\pi)^N\det{\mathbf{K}}}\exp{\left(-\frac{1}{2}\mathbf{v}^T\mathbf{K}^{-1}\mathbf{v}\right)},
\end{equation}
where the covariance matrix $\mathbf{K}$ and waveform difference vector $\mathbf{v}$ are expressed respectively by
\begin{equation}
[\mathbf{K}]_{ij}=k(\boldsymbol{\theta}_i,\boldsymbol{\theta}_j),
\end{equation}
\begin{equation}
[\mathbf{v}]_i=h_\epsilon(\boldsymbol{\theta}_i)-\bar{h}_\epsilon.
\end{equation}
Note that the normalization constant in Eq.~\eqref{eq:PDF} is the square of its usual value for a multivariate Gaussian, due to the two independent channels of the process. Also, $\mathbf{v}$ is a deliberate abuse of notation to cast Eq.~\eqref{eq:PDF} in the familiar Gaussian functional form; its components are itself vectors in $W$ equipped with the inner product \eqref{eq:inner_product}.

The quadratic form in Eq.~\eqref{eq:PDF} may be written as
\begin{equation}
\mathbf{v}^T\mathbf{K}^{-1}\mathbf{v}=\mathrm{tr}\,(\mathbf{K}^{-1}\mathbf{M}),
\end{equation}
with
\begin{equation}\label{eq:overlap_matrix}
[\mathbf{M}]_{ij}=[\mathbf{v}\mathbf{v}^T]_{ij}=\langle h_\epsilon(\boldsymbol{\theta}_i)-\bar{h}_\epsilon|h_\epsilon(\boldsymbol{\theta}_j)-\bar{h}_\epsilon\rangle.
\end{equation}
In \cite{MG2014} and follow-up work, the mean of the process was taken to be the zero vector. Here we use a nonzero but constant mean $\bar{h}_\epsilon$, which is simply chosen to be the mean of the training set of waveform differences $\{h_\epsilon(\boldsymbol{\theta}_i)\,|\,i=1,2,...,N\}$; doing so improves the regression performance at negligible computational cost. We also remove from Eq.~\eqref{eq:overlap_matrix} the factor of $\gamma$ that was introduced in Eq.~(15) of \cite{Chua:2019wgs}. This quantity is defined as the (empirical) ratio between the frequency-averaged power spectral densities of the waveform differences and the detector noise, and was added as a ``fudge factor'' to the base method to prevent the estimate of statistical error from being dominated by the GPR variance when the noise realisation is nonzero. Here, we treat the noise in a more principled way by modifying the definition of the marginalized likelihood in the base method; see Sec.~\ref{subsec:noise likelihood}.


For any new parameter point $\boldsymbol{\theta}$, the enlarged set $\{h_\epsilon(\boldsymbol{\theta}_i),h_\epsilon(\boldsymbol{\theta})\}$ is again normally distributed with mean $\bar{h}_\epsilon$ and the covariance matrix\begin{equation}
\mathbf{K}_*=\left[\begin{array}{cc}\mathbf{K} & \mathbf{k}_*\\\mathbf{k}_*^T & k_{**}\end{array}\right],
\end{equation}
where
\begin{equation}
[\mathbf{k}_*]_i=k(\boldsymbol{\theta}_i,\boldsymbol{\theta}),\quad k_{**}=k(\boldsymbol{\theta},\boldsymbol{\theta}).
\end{equation}
Since $\{h_\epsilon(\boldsymbol{\theta}_i)\}$ is known, the conditional probability distribution of $h_\epsilon(\boldsymbol{\theta})$ given $\{h_\epsilon(\boldsymbol{\theta}_i)\}$ is also Gaussian:
\begin{equation}\label{eq:conditional_PDF}
p(h_\epsilon(\boldsymbol{\theta}))\propto\frac{1}{\sigma^2}\exp{\left(-\frac{1}{2}\frac{\langle h_\epsilon(\boldsymbol{\theta})-\mu|h_\epsilon(\boldsymbol{\theta})-\mu\rangle}{\sigma^2}\right)},
\end{equation}
where $\mu(\boldsymbol{\theta})$ and $\sigma^2(\boldsymbol{\theta})$ are given respectively by
\begin{equation}\label{eq:GPR_mean}
\mu=\mathbf{k}_*^T\mathbf{K}^{-1}\mathbf{v}+\bar{h}_\epsilon,
\end{equation}
\begin{equation}\label{eq:GPR_variance}
\sigma^2=k_{**}-\mathbf{k}_*^T\mathbf{K}^{-1}\mathbf{k}_*.
\end{equation}

Note that $\mathbf{K}^{-1}\mathbf{v}$ in Eq.~\eqref{eq:GPR_mean} and $\mathbf{K}^{-1}$ in Eq.~\eqref{eq:GPR_variance} have nothing to do with $\boldsymbol{\theta}$, and can thus be precomputed. The GPR mean $\mu(\boldsymbol{\theta})$ is essentially an interpolant for $h_\epsilon(\boldsymbol{\theta})$, with associated (squared) error given by the GPR variance $\sigma^2(\boldsymbol{\theta})$. We may thus define a new GPR-informed template model as
\begin{equation}\label{eq:GPR_waveform}
h_\mathrm{GPR}=h_\mathrm{app}-\mu,
\end{equation}
which approximates $h_\mathrm{acc}$ since $h_\mathrm{acc}=h_\mathrm{app}-h_\epsilon$. Eq.~\eqref{eq:conditional_PDF} also provides the prior for $h_\epsilon$ in Eq.~\eqref{eq:marginalisation}, which evaluates to the GPR marginalized likelihood (of the base method):
\begin{equation}\label{eq:marginalised_likelihood}
\mathcal{L}\propto\frac{1}{1+\sigma^2}\exp{\left(-\frac{1}{2}\frac{\langle x-h_\mathrm{GPR}|x-h_\mathrm{GPR}\rangle}{1+\sigma^2}\right)}.
\end{equation}

\subsection{Training the Gaussian process}\label{subsec:GPR training}

The waveform difference model \eqref{eq:GP} is specified by the (fixed) mean of the training set and the covariance function $k$; the latter depends on hyperparameters that are determined by fitting the Gaussian process to the training set. Previous studies in the GW field have demonstrated that the GPR interpolant and the marginalized likelihood function exhibit consistent performance across various common choices for $k$ \cite{MBCG2016}. In this work, we use the squared-exponential covariance function
\begin{equation}\label{eq:SE}
k(\boldsymbol{\theta},\boldsymbol{\theta}')=\sigma_f^2\exp{\left(-\frac{1}{2}\tau^2\right)},
\end{equation}
with
\begin{equation}\label{eq:parameter_distance}
\tau^2=g_{ab}[\boldsymbol{\theta}-\boldsymbol{\theta}']^a[\boldsymbol{\theta}-\boldsymbol{\theta}']^b,
\end{equation}
where the hyperparameters consist only of an overall scale factor $\sigma_f^2$ and the (independent) components $g_{ab}$ of some constant parameter-space metric $\mathbf{g}$ on $\Theta$.

As the size of the training set grows, the covariance matrix $\mathbf{K}$ tends to become ill-conditioned. However, it is common practice to add noise to the training set, which allows for some error in the GPR fit. We transform
\begin{equation}\label{eq:noise}
[\mathbf{K}]_{ij}\to[\mathbf{K}]_{ij}+\sigma_f^2\sigma_n^2\delta_{ij},
\end{equation}
where $\delta_{ij}$ is the Kronecker delta, and the fractional noise variance $\sigma_n^2$ of training-set points is taken to be uniform and fixed. The introduction of noise has the side effect of reducing the condition number of $\mathbf{K}$ for more robust numerical calculations. In this work, we use an empirically determined value of $\sigma_n^2=10^{-2}$ throughout.

The Gaussian process is fit to the training set by maximizing (the logarithm of) Eq.~\eqref{eq:PDF} as a function of the hyperparameters, i.e., the ``hyperlikelihood'' $Z$:
\begin{equation}\label{eq:log-hyperlikelihood}
\ln{Z}=-\frac{1}{2}\mathrm{tr}\,(\mathbf{K}^{-1}\mathbf{M})-\ln{\det{\mathbf{K}}}+\mathrm{const}.
\end{equation}
Part of this maximization may be done analytically, as $\ln{Z}$ with $\mathbf{K}=\sigma_f^2\hat{\mathbf{K}}$ is maximized over $\sigma_f^2$ at
\begin{equation}\label{eq:sigma_f}
\sigma_f^2=\frac{1}{2N}\mathrm{tr}\,(\hat{\mathbf{K}}^{-1}\mathbf{M}).
\end{equation}
Substituting Eq.~\eqref{eq:sigma_f} back into Eq.~\eqref{eq:log-hyperlikelihood}, we may instead maximize the scale-invariant log-hyperlikelihood
\begin{equation}\label{eq:scale-invariant}
\ln{Z}=-N\ln{\mathrm{tr}\,(\mathbf{K}^{-1}\mathbf{M})}-\ln{\det{\mathbf{K}}}+\mathrm{const.}
\end{equation}
over the metric components only, which reduces the dimensionality of the hyperparameter space by one.

\subsection{Fisher-coordinate training grid}\label{subsec:Fisher coordinate}
The waveform derivative $\partial h$ and Fisher information matrix $\Gamma$ are defined respectively as
\begin{equation}\label{eq:derivatives}
[\partial h]_a=\frac{\partial h}{\partial[\boldsymbol{\theta}]^a},\quad[\Gamma]_{ab}=\langle[\partial h]_a|[\partial h]_b\rangle.
\end{equation}
Let $\{(\lambda_i,\hat{\mathbf{v}}_i)\}$ denote the eigensystem of $\Gamma$ for the SNR-normalized waveform difference, $h_\epsilon/\sqrt{\langle h_\epsilon|h_\epsilon\rangle}$, evaluated at some reference parameter point. One can then define a new coordinate system centered on that reference point, by taking the semi-principal axes $\{\lambda_i^{-1/2}\hat{\mathbf{v}}_i\}$ of the associated covariance hyperellipse as basis vectors. A local grid-based training set may be constructed by uniformly placing points on a grid defined by the basis vectors in these ``Fisher coordinates''. Previous research \cite{Chua:2019wgs} has employed this grid-based design.

The main challenge of using a grid-based training set is the issue of scalability in high-dimensional parameter spaces. In this study, we adopt an alternative sampling method and compare it to the traditional grid-based training set in Sec.~\ref{subsec:LHS}. Subsequently, a training set utilizing the alternative sampling method (within a hyperellipse) is employed as our final model in Sec.~\ref{subsec:LISA}.

\section{IMPROVEMENTS \& results}\label{sec:improvements}

As mentioned in Sec.~\ref{sec:level1}, we choose the example of an EMRI signal to showcase our improvements to the base GPR method, and to demonstrate the scalability of our results to the typical length and complexity of EMRI waveforms. Throughout this study, the fiducial model is taken as the augmented analytic kludge \cite{Chua:2016jnd,Chua:2017ujo} with 5PN-adiabatic evolution \cite{Fujita:2020zxe,Isoyama:2021jjd} (5PN AAK), which is publicly available as part of the Fast EMRI Waveforms software package \cite{Katz:2021yft}. This choice is motivated not only by the improved realism of the model relative to previous kludges, but also by its computational efficiency (which provides a tractable fiducial likelihood for comparison with the marginalized likelihood).

To construct an approximate model, we artificially modify the time evolution of the slowly evolving orbital parameters $(p,e,Y)$ (the quasi-Keplerian semi-latus rectum, eccentricity, and cosine of the inclination) by linearly interpolating between 5PN- and 4PN-adiabatic evolution \cite{Sago:2015rpa}:
\begin{equation}
\dot p = (1-c)\dot p_\mathrm{5PN} + c \ \dot p_\mathrm{4PN},
\end{equation}
\begin{equation}
\dot e = (1-c)\dot e_\mathrm{5PN} + c \ \dot e_\mathrm{4PN},
\end{equation}
\begin{equation}
\dot Y = (1-c)\dot Y_\mathrm{5PN} + c \ \dot Y_\mathrm{4PN},
\end{equation}
where $c$ is a tunable quantity that is fixed to 0.0001 in this study. (When $c = 0$, the orbital evolution reduces to 5PN; when $c = 1$, it is equivalent to 4PN.) This construction allows for an approximate model that retains a physically motivated dependence on the EMRI parameters, while producing waveforms that have a controllable overlap with those from the fiducial model.

In this work, an EMRI with redshifted component masses $(\mu,M)_\mathrm{true}=(10^1,10^6)M_\odot$, dimensionless spin parameter $s_\mathrm{true}=0.9$, and initial orbital parameters $(p_0,e_0,Y_0)_\mathrm{true}=(6.97,0.1,0.54)$ is considered as a generic example source. Other source parameters are chosen such that the fiducial signal and the approximate signal have an overlap of 0.84. For simplicity, the long-wavelength approximation for the LISA response \cite{Cutler:1997ta}, $h\equiv(h_\mathrm{I},h_\mathrm{II})$, is used instead of full time-delay-interferometry \cite{Tinto:1999yr,PhysRevD.65.082003,Krolak:2004xp}. The signal is six months long and sampled at 0.2\,Hz, while the source distance is adjusted to yield a high but feasible SNR of 100. The GPR marginalized likelihood Eq.~\eqref{eq:marginalised_likelihood} is used to estimate six source parameters, $(\mu,M,a,p_0,e_0,Y_0)_\mathrm{true}$, assuming all other parameters are known and fixed at their true values. The first application of the base GPR method to EMRIs \cite{Chua:2019wgs} considered only up to a two-month long signal and three estimated parameters.

To decrease the computational expense involved in initializing and evaluating the marginalized likelihood, we employ a band-pass filter as done in \cite{Chua:2019wgs}. This filter is applied to restrict both the data and the templates to the frequency range 3.3--8.3\,mHz, outside of which little signal information is present.

\subsection{Rescaled Fisher coordinates}\label{subsec:re-scaling}
The construction of the training set is based on the Fisher matrix $\Gamma$ for the SNR-normalized waveform difference rather than the Fisher matrix for the accurate waveform, since 
the former more closely approximates the optimal hyperparameter metric (which is SNR-independent); see Sec. III in \cite{Chua:2019wgs} for a more detailed discussion. Thus for SNR values $>1$, the bulk of the likelihood is typically comfortably contained within the span of the training set. However, the covariance hyperellipses associated with both matrices can occasionally still be comparable in scale, especially in the minor directions (corresponding to the largest Fisher eigenvalues). When sampling from the likelihood, the coverage of the training set might thus be insufficient in these directions.

To address this, we adopt a strategy of rescaling the basis vectors of the Fisher coordinates as $\{ \lambda_i^{-1/2}\hat{\mathbf{v}}_i\}\to\{f_i \lambda_i^{-1/2}\hat{\mathbf{v}}_i\}$, so as to boost the training-set coverage in the minor directions. An appropriate choice of $f_i$ would depend on the discrepancy between the covariance hyperellipses for the waveform difference and the accurate waveform (essentially, the former should be rescaled such that it contains the latter). As this discrepancy is model- and signal-specific, such a procedure is necessarily somewhat ad hoc. In this study, we perform the rescaling by hand, with the following empirically determined values: 
\begin{equation}
\begin{aligned}
f_1 & = & 8.0 ,\\
f_2 & = & 4.0 ,\\
f_3 & = & 2.0 ,\\
f_4 & = & 1.0 ,\\
f_5 & = & 0.5 ,\\
f_6 & = & 0.5 ,\\
\end{aligned}
\end{equation}
where the index is sorted in order of decreasing eigenvalue magnitude.

More explicitly, for a training set centered at $(c_1,c_2,c_3,c_4,c_5,c_6)$ in the parameter coordinates, the point $(x_1,x_2,x_3,x_4,x_5,x_6)$ in the rescaled Fisher coordinates corresponds to
\begin{equation}
\begin{bmatrix}
c_1 \\
\vdots \\
c_6 \\
\end{bmatrix}
+
\begin{bmatrix}
f_1 \lambda_1^{-1/2} \hat{\mathbf{v}}_1
& \cdots
& f_6 \lambda_6^{-1/2} \hat{\mathbf{v}}_6
\end{bmatrix}
\begin{bmatrix}
x_1 \\
\vdots \\
x_6 \\
\end{bmatrix}
\end{equation}
in the parameter coordinates. To sum up, our rescaling strategy ensures that the resulting training set sufficiently covers the parameter region of interest, so as to avoid inaccurate inference due to regression error.

\subsection{Fisher-coordinate metric hyperparameters}\label{subsec:training_parameters}

The number of training hyperparameters required to specify the covariance metric in Eq.~\eqref{eq:parameter_distance} scales approximately with the square of the parameter-space dimensionality, which again is a challenge to the scalability of GPR. A common approach to mitigating this in many GPR applications is to use a diagonal metric. Here, we instead use the Fisher matrix $\Gamma$ to place constraints on the metric $g_{ab}$ in Eq.~\eqref{eq:parameter_distance}, since the former is a good approximation to the optimal values for the latter. Specifically, given the unit eigenvectors $\hat{\mathbf{v}}_i$ and eigenvalues $\lambda_i$ of $\Gamma$, we demand that
\begin{equation}
g_{ab} = \left[\begin{matrix} \hat{\mathbf{v}}_1 & \ldots & \hat{\mathbf{v}}_6 \end{matrix}\right]
\begin{bmatrix} \frac{1}{w_1^2} \lambda_1 & & \\ & \ddots & \\ & & \frac{1}{w_6^2} \lambda_6 \end{bmatrix}
\left[\begin{matrix} \hat{\mathbf{v}}_1 \\ \vdots \\ \hat{\mathbf{v}}_6 \end{matrix}\right]^T,
\end{equation}
where now only the $w_i$ are trained hyperparameters. Since $\sigma_f^2$ is fixed by Eq.~\eqref{eq:sigma_f}, the number of hyperparameters is simply the dimensionality of the parameter space. Restricting the covariance function in this way has a negligible impact on regression performance (relative to training the full metric), while significantly reducing the computational cost of training.

\subsection{Latin hypercube sampling}\label{subsec:LHS}

Although a grid-based construction of the training set is simple to implement, it has a couple of important drawbacks. The first one is that the entropy of a grid-based training set is generally lower than that of a more irregularly distributed training set with the same number of points, i.e., it contains less information \cite{shewry1987maximum,sheng2021maximum}. The second drawback of using a grid-based training set in higher dimensions is that a larger number of points will lie in low-likelihood regions, since typical likelihoods have a radial fall-off in density from the maximum-likelihood point.

To address the first drawback, we adopt Latin hypercube sampling (LHS) \cite{mckay2000comparison} as an alternative method of choosing training points. This technique allows a random placement of sample points within a hypercube such that no two samples are aligned (up to a regular partition of the hypercube) in any coordinate direction.
In two dimensions, this is equivalent to the classic problem of placing non-attacking rooks on a chess board.
We adopt a maximum-distance design for generating the Latin hypercube samples, as proposed by \cite{Morris1995a}. This approach aims to maximize the distance between all pairs of samples, while minimizing the number of pairs that are separated by the same distance \cite{Moore1990a}. Thus it prevents highly clustered sample regions, and ensures a more homogeneous distribution of the samples. This variant of LHS is implemented using the Surrogate Modeling Toolbox \cite{SMT2019}.

\begin{figure}[!t]
\includegraphics[width=\columnwidth]{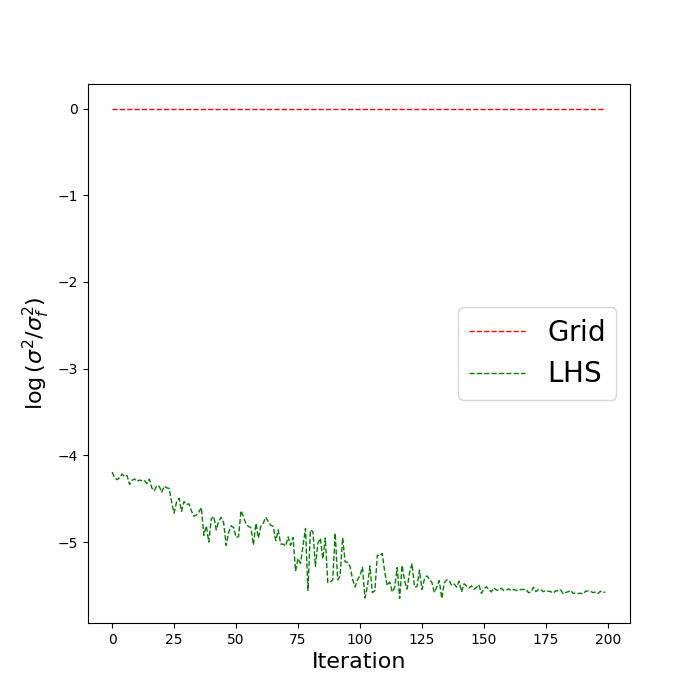}
\caption{\label{fig: evolution_of_sigma} The training evolution of (the logarithm of) the normalised GPR variance $\sigma^2/\sigma_f^2$, evaluated at the center of the training region (the true parameters for the source).}
\end{figure}

In order to compare the performance of an LHS training set against a grid-based training set in GPR, we construct a Fisher-based grid of $N=4^6=4096$ points centered around the signal parameters $\boldsymbol{\theta}_\mathrm{true}$ (such that the grid spans a six-dimensional hypercube of side-length 3 in the rescaled Fisher coordinates), as well as an $N$-sample LHS training set within the same region. The true source parameters are not included in either set of points. From the Gaussian assumption \eqref{eq:PDF}, the entropy of each training set is given by
\begin{equation}
\begin{aligned}
H([h_\epsilon(\boldsymbol{\theta}_i)]) & = -\int \mathrm{D}h_\epsilon\, p([h_\epsilon(\boldsymbol{\theta}_i)])\ln{p([h_\epsilon(\boldsymbol{\theta}_i)])}\\
& =-\mathbb{E}[\ln \mathcal{N}(\bar{h}_\epsilon, \mathbf{K})] \\
& =N(1+\ln{(2 \pi)})+ \ln{\det{\mathbf{K}}}.
\end{aligned}
\end{equation}
Under the same initial hyperparameter values $w_1=\ldots=w_6=1$, the entropy of the grid-based is smaller than that of the LHS training sets by $1651$.
The smaller entropy value for the grid-based training set indicates that it contains less information, which turns out to be insufficient for effective training.

Although the hyperlikelihood for the grid-based training set increases asymptotically towards some optimal value during training, the Gaussian process fails to fit the waveform difference adequately, with the GPR error \eqref{eq:GPR_variance} at most evaluation points of interest taking on its maximal value ($\sigma_f^2$ in Eq.~\eqref{eq:sigma_f}). This is illustrated by the top plot in Fig.~\ref{fig: evolution_of_sigma}, which shows how the GPR variance (normalised by $\sigma_f^2$) at the true signal parameters fails to improve as training proceeds. In contrast, the same plot for the LHS training set tends toward a minimal value $\ll1$, indicating that the set is more optimal for regression while having the same span and number of points as the grid-based training set.

\begin{figure*}[!tbh]
\includegraphics[width=\textwidth]{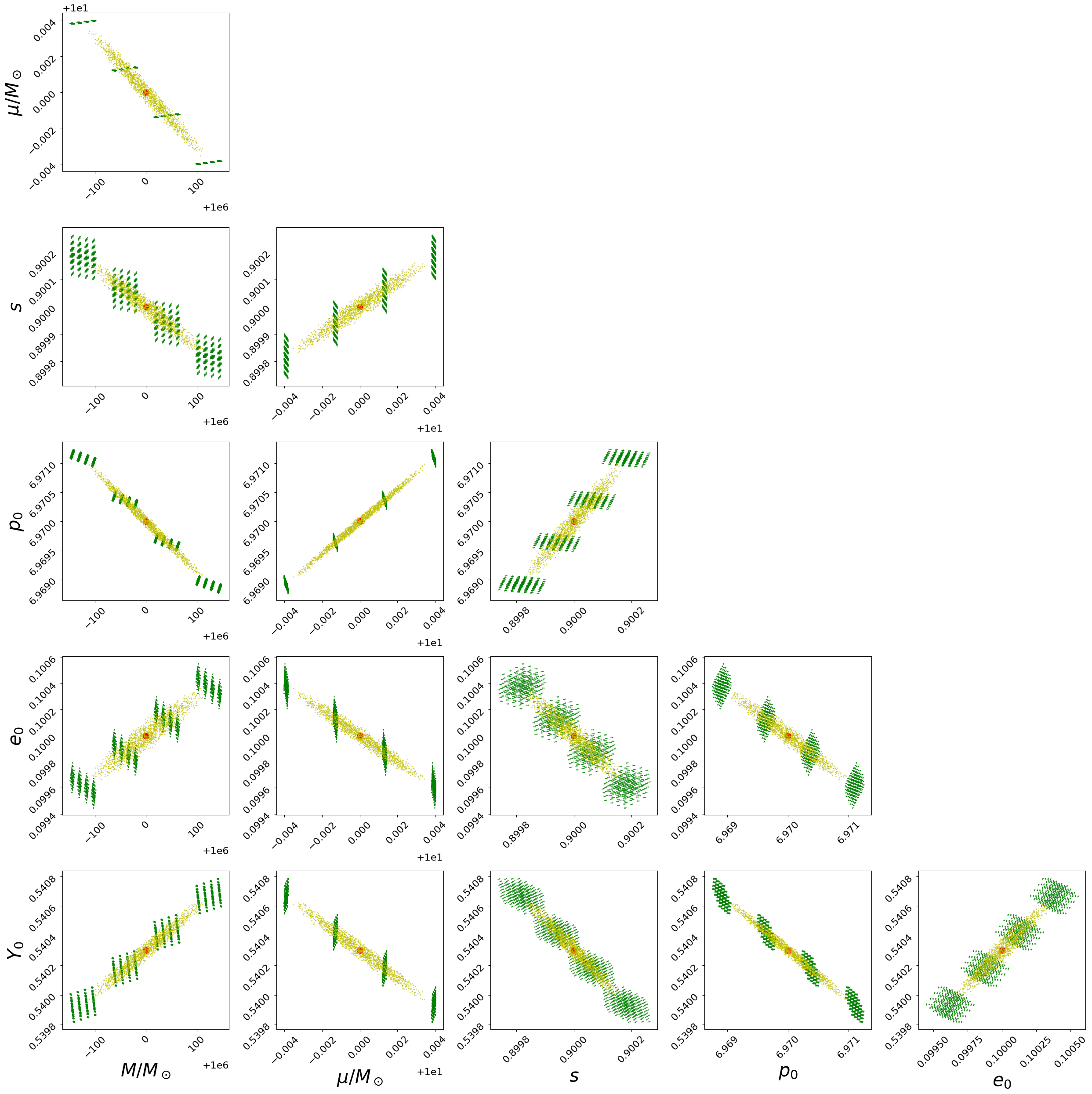}
\caption{\label{fig: trainingset_points} Visualization of the grid-based training set in Sec.~\ref{subsec:LHS} (4096 green points) and the hyperspherical truncation of the LHS training set with the same span (1611 yellow points), both centered on the true parameters for the source (red point).
}
\end{figure*}

As for the second drawback of using a grid-based training set, the larger relative volume contained in the ``corners'' of the hypercube leads to a larger proportion of uninformative points in the set, which adds unnecessary computational cost to both the training and evaluation of the GPR model. To make this intuitive, consider a hypersphere (representing the bulk of the likelihood density) that is inscribed in a hypercube (representing the span of the training grid), in $d$ dimensions. When $d=2$, the volume outside the hypersphere is $21\%$ of the hypercube volume; this rises to $\geq92\%$ when $d\geq6$.

To address this drawback, we implement a further hyperspherical truncation of the LHS training set in our final model (used in Sec.~\ref{subsec:LISA}). In the rescaled Fisher coordinates, the covariance hyperellipse associated with the Fisher matrix is a hypersphere; we simply enlarge this such that it is inscribed in the hypercube spanned by the grid, and then remove all LHS points lying outside the enlarged hypersphere. In this way, the number of model evaluations in low-likelihood regions is greatly reduced. Both the grid-based training set and the truncated LHS training set are compared visually in Fig.~\ref{fig: trainingset_points}.

\subsection{Computational acceleration of training}\label{subsec:efficiency}

When training the Gaussian process, the cost of evaluating the hyperlikelihood is dominated by the calculation of $\mathbf{K}^{-1}\mathbf{M}$ at each iteration, especially for large $N$. The most efficient way of computing this quantity is then: i) to solve the linear systems of equations $\mathbf{K}\mathbf{X}=\mathbf{M}$ for $\mathbf{X}$ (instead of inverting $\mathbf{K}$), and ii) to parallelize the calculation by performing it on a GPU. In previous work \cite{MBCG2016,Chua:2019wgs}, the gains from this approach were marginal even for the largest considered training sets with $N\sim10^2$. In this work, where $N\gtrsim10^3$, it becomes essential. We use the conjugate gradient method \cite{hestenes1952methods} to solve for the roots of $\mathbf{K}\mathbf{X}-\mathbf{M}$; this is an iterative technique that is better suited to large $N$ than previously employed methods such as Cholesky decomposition. For a training set containing 4096 points, a single training iteration typically takes around 3.5 seconds when evaluated on a GPU, and around 200 iterations in total to converge.

GPUs can also be used to accelerate evaluation of the trained model (now with fixed $\mathbf{K}$, such that $\mathbf{K}^{-1}\mathbf{v}$ in Eq.~\eqref{eq:GPR_mean} and $\mathbf{K}^{-1}$ in Eq.~\eqref{eq:GPR_variance} can be precomputed). A single evaluation of the GPR mean and variance takes around 0.01s on a GPU, as compared to around 0.1s on a CPU. However, note that the marginalized likelihood requires evaluation of the approximate waveform, which must also be accelerated in order to gain the full benefit from the Gaussian-process component of the model. In this work, as the cost of waveform generation is the computational bottleneck in the case of EMRIs, we do not implement the sampling of likelihoods on a GPU.

\subsection{Marginalized likelihood for nonzero noise}\label{subsec:noise likelihood}

From standard noise properties, the expectation of the logarithm of Eq.~\eqref{eq:marginalised_likelihood} is given (up to a constant) by
\begin{equation}\label{eq:E_ln_marginalised_likelihood}
\text{E}[\ln{\mathcal{L}}] = -\frac{1}{2}\frac{\langle s-h_\mathrm{GPR}|s-h_\mathrm{GPR}\rangle+\mathcal{N}}{1+\sigma^2}+\ln{\left(\frac{1}{1+\sigma^2}\right)},
\end{equation}
where $\mathcal{N} \coloneqq \text{E}[\langle n|n\rangle]$ is the expected noise power. It is the interplay between $\mathcal{N}\neq0$ and the size/variation of $\sigma^2$ over the signal space that is generally problematic for the practical application of the GPR marginalized likelihood with nonzero noise. Essentially, the profile of the likelihood becomes driven by the variation of $\sigma^2$ if $\mathcal{N}$ is too large, and it can even have a narrowed credible region that excludes the true parameters to high significance.

This issue was recognised and addressed in \cite{Chua:2019wgs}, although not explicitly described in that paper. There, $\sigma^2$ was reduced by an overall factor of $\gamma\ll1$, with the value of $\gamma$ chosen empirically as the ratio between the typical power of the waveform differences and the expected power of the detector noise. This works simply because the GPR likelihood approaches the accurate likelihood as $\gamma\to0$, but it is rather ad hoc and does not generally yield broadened credible regions in the former.

We will instead redesign the GPR likelihood in a way that aims to recover, for general noise realizations, its behaviour when $\mathcal{N}=0$ (while reducing to the accurate likelihood as $\sigma^2\to0$). It is straightforward to achieve this when the likelihood is approximated by Eq.~\eqref{eq:E_ln_marginalised_likelihood}; one such solution is simply
\begin{equation}\label{eq:new_marginalised_likelihood}
\mathcal{L}\propto\frac{1}{1+\sigma^2}\exp{\left(-\frac{1}{2}\frac{\langle x-h_\mathrm{GPR}|x-h_\mathrm{GPR}\rangle+\mathcal{N}\sigma^2}{1+\sigma^2}\right)}.
\end{equation}
Here, $\mathcal{N}$ needs to be estimated accurately because the noise-corrected likelihood can still be quite sensitive to any residual noise power near the maximum-likelihood point. We propose an iterative method for estimating $\mathcal{N}$ (and for refining inference) in the following section.

\subsection{Iterative inference}\label{subsec:iterative GPR}

When applying the GPR method to realistic inference, the starting point is an estimate of the true source parameters $\boldsymbol{\theta}_\mathrm{true}$, so as to construct the training set in its local vicinity. This estimate is most naturally obtained through maximum likelihood or maximum a posteriori estimation with the approximate waveform model (since the accurate model is assumed to be computationally intractable); we denote it by $\boldsymbol{\theta}_\mathrm{app}$. At this stage, a first estimate of $\mathcal{N}$ is given by
\begin{equation}
    \mathcal{N}_\mathrm{1stGPR}=\langle x-h_\mathrm{app}(\boldsymbol{\theta}_\mathrm{app})|x-h_\mathrm{app}(\boldsymbol{\theta}_\mathrm{app})\rangle.
\end{equation}
Together with the training set centered on $\boldsymbol{\theta}_\mathrm{app}$ and the Gaussian process that is trained on this set, inference can then be performed using the noise-corrected GPR likelihood \eqref{eq:new_marginalised_likelihood}, which we denote by $\mathcal{L}_\mathrm{1stGPR}$.

Computation of the posterior under the first GPR model yields a first maximum a posteriori estimate, which we denote by $\boldsymbol{\theta}_\mathrm{1stGPR}$ ($\sigma^2$ taken as 0 in this optimization). Depending on the distance between $\boldsymbol{\theta}_\mathrm{1stGPR}$ and $\boldsymbol{\theta}_\mathrm{true}$ (equivalently, the error in the approximate model), the first GPR posterior may not be a sufficiently faithful approximation to the accurate posterior. In this case, another iteration of inference may be performed by constructing a second training set centered on $\boldsymbol{\theta}_\mathrm{1stGPR}$, retraining the Gaussian process, and re-estimating the noise as
\begin{equation}\label{eq:N}
\mathcal{N}_\mathrm{2ndGPR} = \langle x-h_\mathrm{acc}(\boldsymbol{\theta}_\mathrm{1stGPR})|x-h_\mathrm{acc}(\boldsymbol{\theta}_\mathrm{1stGPR}) \rangle .
\end{equation}
Here we have used the accurate waveform model to compute the second noise estimate, although we could also have used $h_\mathrm{1stGPR}$ instead. In realistic scenarios, such an iterative usage of the GPR method for high-precision inference is unlikely to take more than two iterations; if the error in the approximate model is so large as to require this, a more prudent approach in the first place would be to improve the approximate model or construct better fits to the accurate model.

\subsection{Results}\label{subsec:LISA}
In this subsection, we present illustrative results for the GPR method with all of the proposed improvements in Secs \ref{subsec:re-scaling}--\ref{subsec:noise likelihood}, when applied to our example EMRI signal with simulated LISA noise \cite{Robson:2018ifk}. We assume a flat prior in a suitably bounded region of parameter space, and employ the Markov chain Monte Carlo sampler emcee \cite{H1970,Foreman-Mackey:2012any} to draw samples from the standard approximate likelihood $L_\mathrm{app}$, the standard accurate likelihood $L_\mathrm{acc}$, and the GPR marginalized likelihoods $\mathcal{L}$. Note that the accurate likelihood is assumed to be unavailable in the actual scenarios to which the GPR likelihood might be applied, but we include it here to showcase the efficacy of the method.

\begin{figure*}[!htb]
\includegraphics[width=\textwidth]{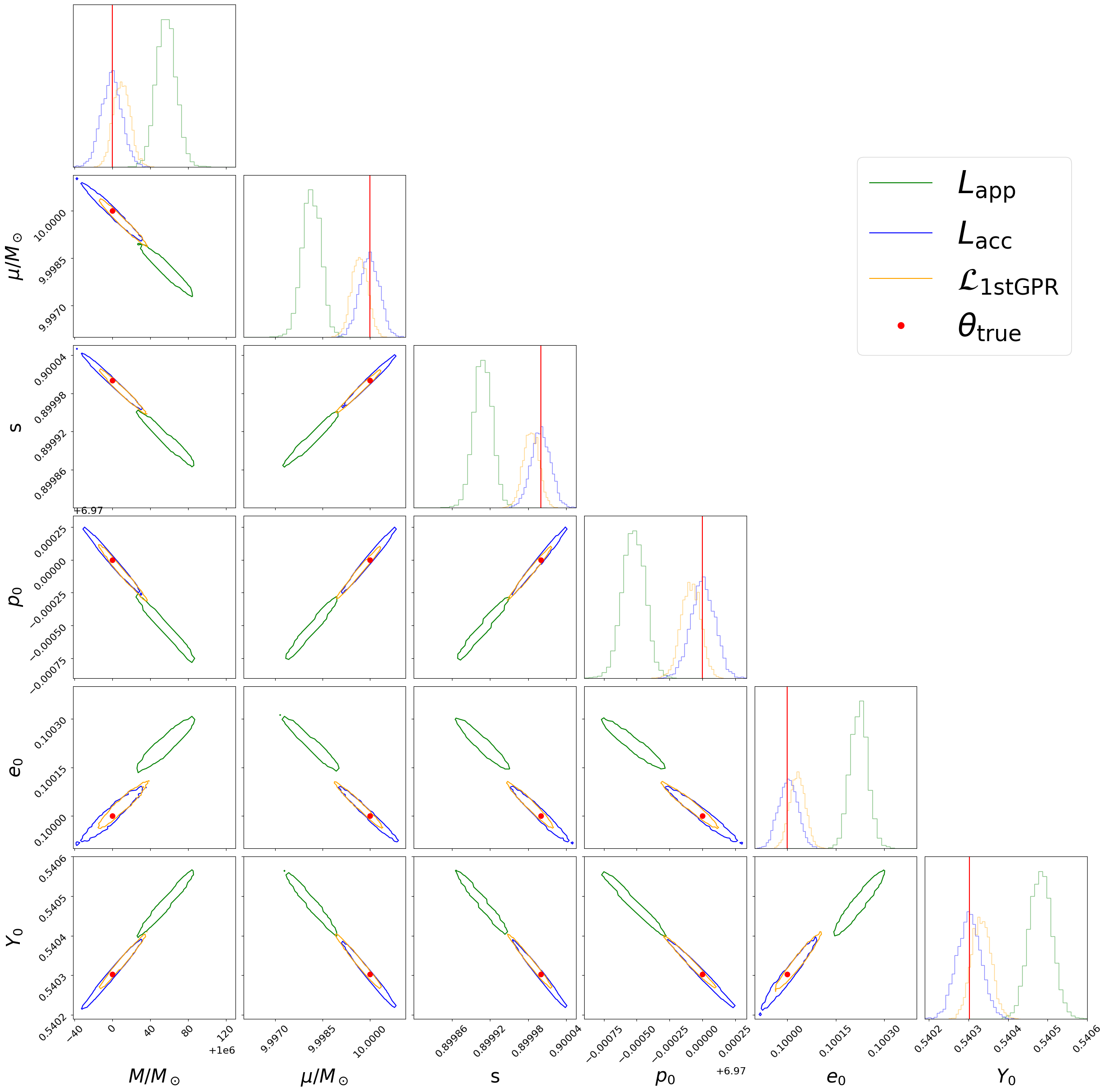}
\caption{\label{fig:noise_1stGPR_MCMC_points} Projected one- and two-dimensional plots comparing the accurate and approximate likelihoods to the first noise-corrected marginalized likelihood (with a truncated LHS training set that is centered at the maximum likelihood estimate of the approximate waveform). The shown contour level is 3-sigma. The true source parameters are indicated by red dots.}
\end{figure*}

\begin{figure*}[!htb]
\includegraphics[width=\textwidth]{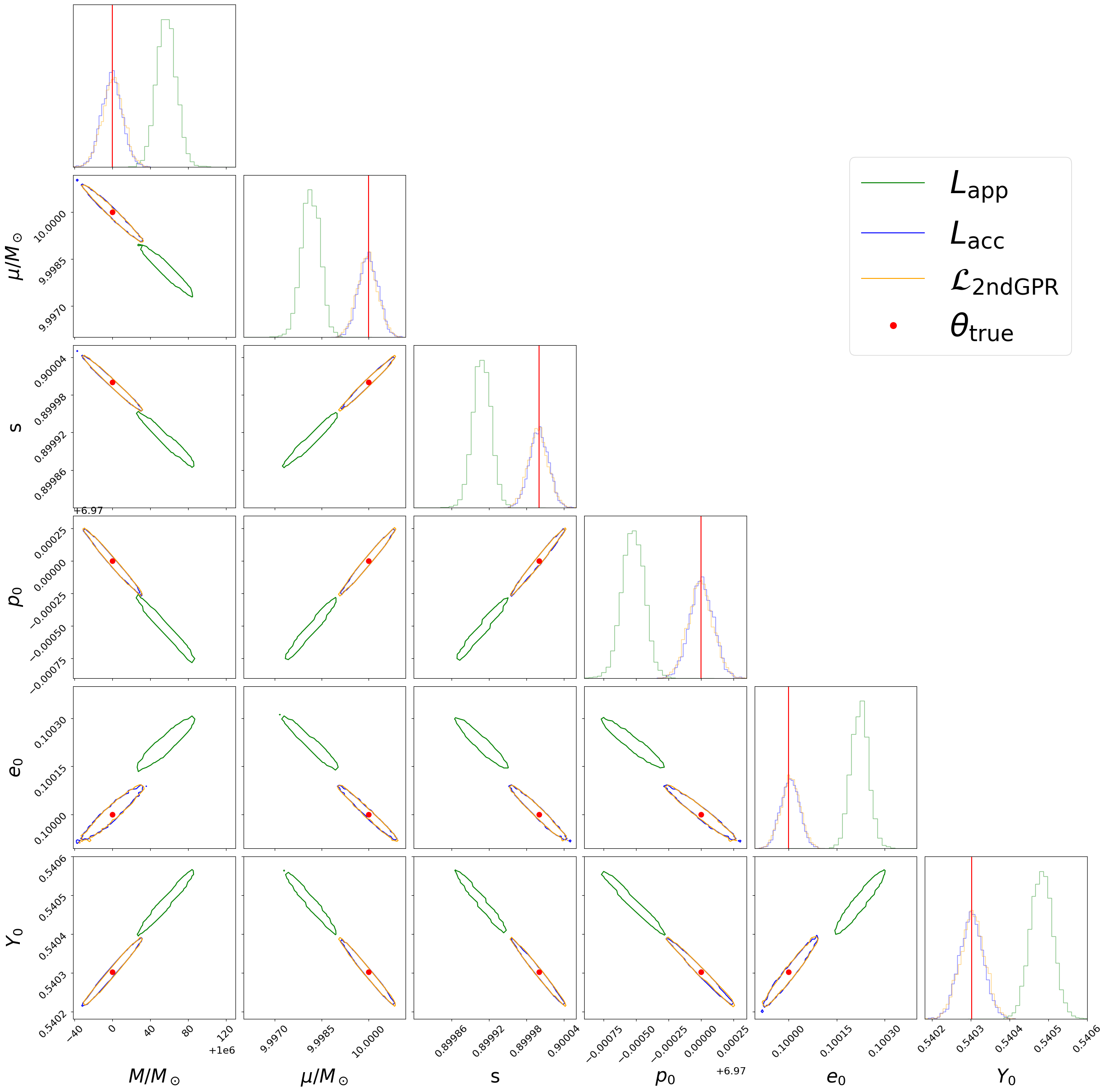}
\caption{\label{fig:noise_2ndGPR_MCMC_points} Projected one- and two-dimensional plots comparing the accurate and approximate likelihoods to the second noise-corrected marginalized likelihood (with a truncated LHS training set that is centered at the maximum likelihood estimate of the first GPR model). The shown contour level is 3-sigma. The true source parameters are indicated by red dots.}
\end{figure*}

Fig.~\ref{fig:noise_1stGPR_MCMC_points} displays the GPR likelihood that is computed in the first iteration described in Sec.~\ref{subsec:iterative GPR}, along with the standard accurate and approximate likelihoods for comparison. The approximate likelihood excludes the true source parameters at more than 3-sigma significance. On the other hand, the first GPR likelihood provides a decent (but still slightly shifted) approximation to the accurate likelihood, and agrees with the true parameters to within 3 sigma.
Finally, results from a second iteration are presented in Fig. \ref{fig:noise_2ndGPR_MCMC_points}, where the second GPR likelihood is seen to be almost perfectly consistent with the accurate likelihood.

\section{Conclusion}\label{sec:conclusion}
This work improves the scalability of the GPR marginalized-likelihood scheme for high-precision GW inference \cite{MG2014,MBCG2016,Chua:2019wgs}, thus extending its potential application to higher-dimensional parameter spaces and longer signal durations (six intrinsic parameters and six-month long signals, in our EMRI example). Several significant modifications have been made to the base GPR method that was developed in previous work.

In Secs \ref{subsec:re-scaling}--\ref{subsec:efficiency}, various improvements to the training of the Gaussian process are described. These are: (i) a rescaling of the Fisher-informed training set that is better adapted to highly correlated parameters; (ii) a Fisher-informed constraint on the covariance metric such that the number of hyperparameters scales linearly rather than quadratically with the parameter-space dimensionality; (iii) the use of LHS and (Fisher-informed) hyperspherical truncation to construct the training set; and (iv) computationally efficient training through the use of the conjugate gradient method and GPU acceleration. These modifications boost the scalability of the GPR method by significantly reducing the required density of the training set in a given region of interest (such that it grows sub-exponentially with the parameter-space dimensionality), and by greatly accelerating the training process as well.

Secs \ref{subsec:noise likelihood} and \ref{subsec:iterative GPR} describe improvements to the marginalized-likelihood method itself. We make a crucial redefinition of the marginalized likelihood in order to render it usable in realistic inference scenarios with nonzero detector noise; this is done via an estimation of the noise power by computing the data--template residual at the maximum likelihood parameters. We also propose an iterative approach to using the GPR method, where the training set and noise estimate are refined through (a single) repetition. Finally, in Sec.~\ref{subsec:LISA}, we implement all of the above improvements to perform inference on an example EMRI signal with simulated LISA noise, and to demonstrate the viability of the GPR marginalized-likelihood method for more realistic GW applications.

\begin{acknowledgments}
ML would like to express his gratitude to Prof. Yiming Hu and Prof. Jiandong Zhang for their insightful discussions and valuable feedback on our research. ML is also grateful to the NUS Research Scholarship for its financial support. AJKC thanks Christopher Moore for helpful comments on the manuscript, and acknowledges previous support from the NASA LISA Preparatory Science grant 20-LPS20-0005.
\end{acknowledgments}

\nocite{*}

\bibliography{apssamp}
\bibliographystyle{apsrev4-1}

  

\end{document}